\documentclass[11pt,a4paper]{article}
\usepackage{amsmath,amssymb}
\usepackage{epsfig,graphicx}

\newcommand\ap[3]    {
                {{\it Ann.\ Phys.\ (NY) }{\bf #1} (#2) #3}}
\newcommand\npb[3]    {
                {{\it Nucl.\ Phys.\ }{\bf B #1} (#2) #3}}
\newcommand\plb[3]   {
                {{\it Phys.\ Lett.\ }{\bf B #1} (#2) #3}}

\newcommand\npps[3]  {
                {{\it Nucl.\ Phys.\ }{\bf #1} {\it(Proc.\ Suppl.)} (#2) #3}}
\newcommand\jetpl[3]  {
                {{\it Sov.\ Phys.\ JETP Lett.\ }{\bf #1} (#2) #3}}
\newcommand\sjnp[3]  {
                {{\it Sov.\ J.\ Nucl.\ Phys.\ }{\bf #1} (#2) #3}}

\newcommand{\hepph}[1]{{\tt hep-ph/#1}}

\let \red=\bf
\let\blue=\relax

\renewcommand\t{\theta}
\renewcommand\l{\lambda}
\renewcommand\a{\alpha}
\renewcommand\b{\beta}

\renewcommand\P{{\hat P}}
\newcommand\e{\ensuremath{\epsilon}}

\newcommand\R[1]{\ensuremath{\mathbf{#1}}}
\newcommand\Rb[1]{\ensuremath{\overline{\mathbf{#1}}}}
\newcommand\tb{\ensuremath{\mathop{\rm tan}\beta}}

\newcommand\eV{\mbox{eV}}
\newcommand\MeV{\mbox{MeV}}
\newcommand\GeV{\mbox{GeV}}
\newcommand\TeV{\mbox{TeV}}

\newcommand\Md{\ensuremath{m_d}}
\newcommand\Ms{\ensuremath{m_s}}

\newcommand\be{\begin{equation}}
\newcommand\ee{\end{equation}}
\newcommand\bea{\begin{eqnarray}}
\newcommand\eea{\end{eqnarray}}
\newcommand\ba{\begin{array}}
\newcommand\ea{\end{array}}
\newcommand\bma{\begin{array}{ccc}}
\newcommand\ema{\end{array}}
\newcommand\matr[1]{\left(\bma#1\ema\right)}
\newcommand\0{\nonumber}

\begin{document}

\title{\bf Fermion Masses at intermediate $\tb$: Unification of Yukawa
Determinants}

\author{F. Nesti\footnote{{\bf e-mail}: fabrizio.nesti@aquila.infn.it}\\
\small{\em Universit\`a dell'Aquila, INFN sez.\ dell'Aquila} \\
\small{\em v. Vetoio 1, 67010, L'Aquila, Italy}
}
\date{}
\maketitle

\begin{abstract}
In the context of the Grand Unified MSSM, we investigate the fermion
mass matrices at GUT scale. We note that from the experimental mass
pattern the determinants of the Yukawa matrices at this scale can be
unified with good precision. Taking the unification o determinants as
an hypothesis, it gives two model independent predictions that in the
MSSM turns out to determine an appropriate value for the product
$m_dm_s$ and $\tb\simeq7$--$10$ in the favored range. We then review
a predictive model of SU(3) flavour in the context of supersymmetric
SO(10) that nicely implements this mechanism, while explaining all
fermion masses and mixings at 1$\sigma$ level, including neutrino
data.
\end{abstract}

\bigskip
\bigskip
\bigskip

\section{Introduction}

Understanding the physics that underlies the measured pattern of
fermion masses and mixings is still an open problem. While there is
little clue on the way to follow, it is certain that one will need a
framework beyond the Standard Model. We approach here the problem in
the interesting setup of SO(10) Supersymmetric Grand Unification (SUSY
GUT)~\cite{so10} with an horizontal SU(3) flavour
symmetry~\cite{SU3H}.  The reason for this choice is of course
that all the fermions of a Standard Model family plus the right-handed
neutrino fit in a basic multiplet of SO(10), the $\R{16}$
representation, and that the SU(3) flavour symmetry is the largest
allowed (actually U(3)) by gauge interactions.  SUSY provides a
one-scale unification of the gauge couplings, in addition to benefits
like the stability of the (large) hierarchy, and others.

While in the low energy world both the gauge SO(10) and flavour SU(3)
symmetries are broken, in this framework the Yukawa matrices are
treated in an unified way, so that they will show the remnants of both
these breakings.

With this in mind we approach the problem by studying the mass
matrices at the Grand-Unification scale $M_{GUT}\simeq2\cdot
10^{16}\,\GeV$. We consider first the charged-fermions Yukawa
couplings renormalized at this common scale assuming MSSM running
(recalled in appendix~\ref{app:RG}). We plot them in
figure~\ref{fig:abc}, showing also the ranges resulting from
experimental uncertainties, and illustrating the three relevant cases
of small, moderate and large~$\tb$.

\begin{figure}[t]%
  \epsfig{file=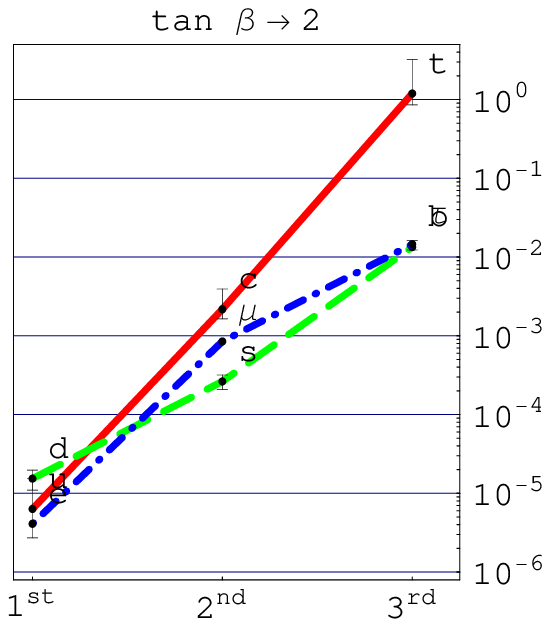,height=18em,width=13em}
  \epsfig{file=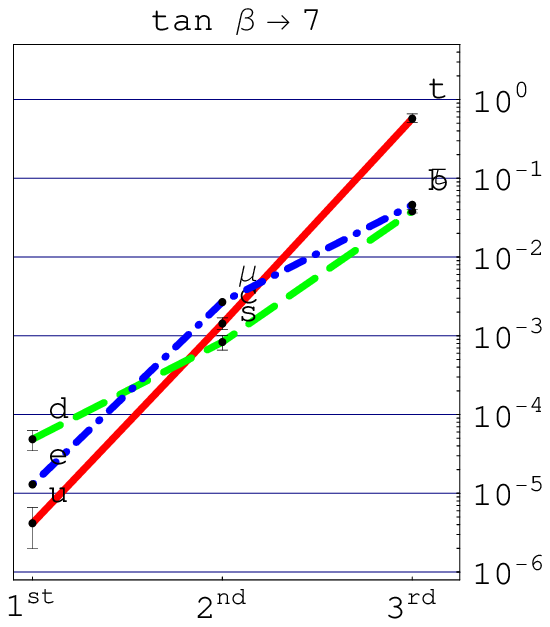,height=18em,width=13em}
  \epsfig{file=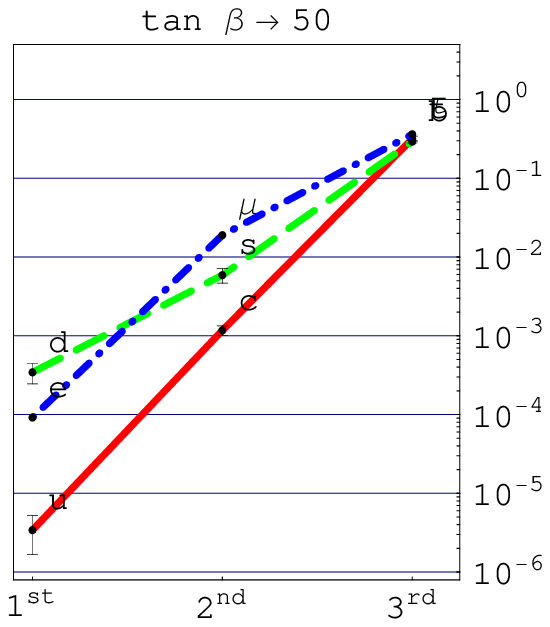,height=18em,width=13em}%
  \caption{Yukawa couplings at GUT scale calculated from experimental
  charged fermions masses, assuming MSSM running and including
  uncertainties also from $\l_t$ renormalization. U, D, E are
  respectively red (continuous), green (dashed) and blue
  (dot-dashed).\label{fig:abc}}
\end{figure}

The first observation that one can make in figure~\ref{fig:abc} is that
the Yukawa couplings show clear hierarchic patterns: indeed if it were
not for some deviations, the picture would be quite simple:
$\l_f\sim\e_f^2:\e_f:1$, with small parameters $\e_f$, for $f=u,d,e$.
One can estimate the hierarchy parameters as $\e_u\sim1/500$ in the
``up'' sector, while in the ``down'' and ``charged leptons'' sectors
the hierarchies are very similar, $\e_{d,e}\simeq 1/50$, reaching the
well known observation that $\e_u\ll\e_e\simeq\e_d$.

About the deviations, it is evident that the ``up'' sector (U) has an
almost exact hierarchy, while the ``down'' and ``charged leptons''
sectors (D, E) deviate sensibly.  These deviations can be thought to
belong mainly to the first two generations, so that the picture that
emerges is the following:%
\vspace*{-2ex}
\bea
\label{eq:up}
\l_u\ :\ \l_c  \ :\ \l_t   &\ \sim\ &\ \e_u^2\ \ \ \ :\ \ \ \e_u\ \ \
:\ \ 1\\[1.1ex]
\l_d\ :\ \l_s  \ :\ \l_b   &\ \sim\ &\frac{1}{A_d}\,\e_d^2\ :\ A_d\,\e_d\
:\ \ 1\\
\l_e\ :\ \l_\mu\ :\ \l_\tau&\ \sim\ &\frac{1}{A_e}\,\e_e^2\ :\ A_e\,\e_e\
:\ \ 1\,.
\eea
The factors $A_e$, $A_d$ that parametrize the deviations can also be
estimated as $A_e\simeq2.5$ and $A_d\simeq0.5$.

Other noticeable facts that can be deduced qualitatively from
figure~\ref{fig:abc} are: 1) $b$-$\tau$ unification,
$\l_b\simeq\l_\tau$, that works in all $\tb$ regime; 2) at large
$\tb\sim50$ also ``top'' is unified, $\l_t\simeq\l_b\simeq\l_\tau$; 3)
at very small $\tb\sim2$ the first family is approximately unified, as
is the second family at intermediate $\tb\sim7$.  Aspects 1) and 2)
were analyzed deeply, see e.g.~\cite{BBO}, while very small $\tb$ was
suggested as a possible \emph{inverted hierarchy}
unification~\cite{br,z}. Here we will focus on the unification of the
second family, that works at intermediate $\tb$ and leads to an
interesting scenario.

As far as neutrinos are concerned, we have no direct information on
their Yukawa couplings: even though we now know that they have a mass
and we know their hierarchy and mixing angles, the right-handed
neutrinos, that are automatically present in SO(10), may have a
separate Majorana mass matrix. Therefore one should not expect
neutrinos to follow the same scheme of the charged fermions, and will
be considered later.

\section{Unification of determinants}
\label{sec:detunif}

A step beyond these observations can be made by using together two
known ``vertical'' relations between the D and E sectors, namely that
$\l_d\l_s\simeq\l_e\l_\mu$ and $\l_b\simeq\l_\tau$.%
\footnote{The first holds precisely but with 40\% experimental
uncertainty from $m_dm_s$, the second is exact to 10\%, with $m_b$
known to 5\% accuracy.}
Together, these two relations lead to the conclusion that for the D, E
sectors one has
\be
\l_e\,\l_\mu\,\l_\tau \simeq \l_d\,\l_s\,\l_b\,.
\ee
In words, since what appear here are the eigenvalues of the Yukawa
matrices, this simple relation suggest that in the D and E sectors the
determinants of the Yukawa matrices are unified at GUT scale.


It is natural to try to extend the previous argument also to the U
sector.  In doing so, in the MSSM, we encounter the presence of the
$\tb$ parameter, that measures the U-D Higgs VEV orientation and
shifts the U sector Yukawa eigenvalues with respect to the D, E ones;
this can be seen in the three cases of figure~\ref{fig:abc}. 

Since $\tb$ is very loosely bounded, until SUSY is discovered, we may
proceed in the inverse direction: we may assume the unification of the
Yukawa determinants, as if it was motivated by some symmetry
principle, and look for its phenomenological consequences. 

We find the first result in terms of $\tb$: if we impose
\be
\label{eq:yukunif}
\l_e\,\l_\mu\,\l_\tau =\l_d\,\l_s\,\l_b=\l_u\,\l_c\,\l_t\,,
\ee
it follows that among the three regimes of small, moderate or large
$\tb$, only the middle one is selected.  Therefore unification of
determinants gives a prediction of $\tb\simeq7$.  This qualitative
conclusion may be checked in figure~\ref{fig:abc}, but we will describe
below a precise test of it. It is interesting to note that this value
of $\tb$ lies in the currently favored range~\cite{higgs}.

Then, still reasoning in the inverse direction, after choosing the
appropriate $\tb\simeq7$ that guarantees the unification of the U with
the D, E determinants, we have that all Yukawa eigenvalues are built
around the common Yukawa scale $\l$ defined precisely by
$\l^3=\l_e\,\l_\mu\,\l_\tau$, since the ``lepton'' Yukawa couplings
are known with best accuracy. Now we note that this scale acts as a
``pivot'' for all the hierarchies: since fermions are hierarchic, the
first family fermions get lighter when the third family ones are
heavier.  The middle scale can be approximately identified with
$\l_c$, because the U Yukawa couplings are hierarchic to a very good
approximation.

In the U sector we find the first and simpler example: the more
$\l_t$ is large the lower lies $\l_u$:
\be
\l_u=\frac{\l^2}{\l_t}\,,\qquad\mbox{with }\l^3\simeq\l_e\l_\mu\l_\tau\,.
\ee
One can think of this mechanism as a kind of seesaw between different
flavours, i.e.\ a {\em flavour seesaw}.\footnote{Of course until a
realization in flavour space is given, the flavour-seesaw mechanism
described here has not the commonly assumed meaning of 'mixing with
heavy decoupled states', but it has the original meaning of the
seesaw: a quantity going down whenever an other goes up\ldots}

In the D, E sectors similarly, since $\l_\tau$, $\l_b$ are large then
$\l_e\l_\mu$ and $\l_d\l_s$ will be small. Moreover, a similar
phenomenon takes place: using bottom-tau unification, $\l_b\simeq
\l_\tau$, one has that $\l_e\l_\mu\simeq\l_d\l_s=\l^3/\l_{b,\tau}$.
This relation says that in the D and E sectors, thanks to $b$-$\tau$
unification, in the first two generations there is a smaller seesaw:
the deviations of $\l_\mu$ and $\l_e$ are in opposite directions with
respect to exact hierarchy, and the same happens for the deviations of $\l_d$
and $\l_s$.

\medskip

Of course this mechanisms can not explain why $\l_t$ or $\l_b$ are
higher than the common $\l$, like the celebrated seesaw mechanism in
the neutrino sector can not explain the high scale of the RH neutrino
Majorana mass. The unification of determinants only provides the two
relations~(\ref{eq:yukunif}) that should be accompanied by a complete
``theory of fermion masses'' to generate the actual hierarchies
between yukawas, and also their deviations.

We will describe below one such a model of Flavour SUSY GUT that
includes a natural realization of these ideas, but before we will
analyze the model independent predictions that strictly follow from
the unification of Yukawa determinants~(\ref{eq:yukunif}) alone.

\medskip

The conditions~(\ref{eq:yukunif}) give two testable relations between
quarks and lepton masses. In particular, running down to low energy
(see appendix \ref{app:RG}):
\bea
\label{eq:dspred}
m_d m_s&=&\frac{m_e m_\mu m_\tau}{m_b}\, \frac{R_d^3
  \eta_b\eta_{uds}^2}{R_e^3 \eta_{e\mu}^2\eta_\tau} B_t\\[1ex]
\label{eq:upred}
m_u &=&\frac{m_e m_\mu m_\tau}{m_c m_t}\, \frac{R_u^3
  \eta_c\eta_{uds}}{R_e^3 \eta_{e\mu}^2\eta_\tau} (\tb)^3 B_t^{12}
\eea
and these translate into the following predictions for $m_dm_s$ and $m_u$:
\bea
\label{eq:predmdms}
m_dm_s&=&880 \,\MeV^2\pm 2\%_{m_t}\pm4\%_{m_b}\pm25\%_{\a_3}\label{eq:ds}\\[1ex]
\label{eq:predmu}
m_u&=&\bigg(\frac\tb7\bigg)^3 1.95\,\MeV\pm20\%_{m_t}\pm10\%_{m_c}\pm15\%_{\a_3}\,.\label{eq:mu}
\eea
Using a mean value of 19.5 for the known ratio $m_s/m_d$, the first equation
gives the central values
\be
 m_d=6.7\,\MeV\,,\qquad m_s=131\,\MeV\,,
\ee
that fit well the data, with $m_s$ a bit high, within 1$\sigma$ of the
experimental range. This means that from experimental data the D, E
determinants are unified at GUT scale within 1-$\sigma$. Lowering
$\a_3|_Z$ helps this unification, so does raising the SUSY scale.

The prediction for $m_u$ carries a dependence on \tb, that can be used
to derive a prediction for it, as a function of $m_u$ itself:
\be
\label{eq:predtb1}
\tb=7.3\,\pm1.3_{m_u}\,\pm0.5_{m_t}\,\pm0.2_{m_c}\,,
\ee
where we used the 1$\sigma$ range for $m_u=(3\pm
1.5)\,\MeV$~\cite{pdg}. This prediction lies exactly in the favored
range determined from the recent analysis in various MSSM
scenarios~\cite{pdg,higgs}.

An alternative prediction for $\tb$ may be derived by combining the two
above and assuming for $m_u/m_d$ and $m_s$ their experimental
ranges. One gets:
\bea
\label{eq:predtb2}
\tb&=&\bigg(\frac{m_u}{m_d}\,\frac{m_cm_t}{m_sm_b}\bigg)^{1/3}
      \frac{R_u}{R_d}\bigg(\frac{\eta_b\eta_{uds}}{\eta_c}\bigg)^{1/3}
      B_t^{-11/3}\0\\[1ex]
&=&10\,\pm1.3_{m_u/m_d}\,\pm1_{m_s}\,\pm 1_{m_t}\,\pm0.5_{\a_3}\,.
\eea
This prediction tells what is the value of \tb\ once a model has been
solved and $m_u/m_d$, $m_s$ and $m_t$ are determined.  

We can note however that the dependence on $\a_3|_Z$ cancels almost
perfectly in both the ratios involving the $R$'s and the $\eta$'s, and
enters only through the $B_t$ dependence ($<\!4\%$). This is a
remarkable fact, considering that for individual masses the
uncertainty in $\a_s|_Z$ generates the dominant errors in their
predictions.

In addition, and equally remarkably, the even larger dependence on the
supersymmetry breaking scale also cancels almost exactly in the ratio
$R_u/R_d$, and enters just through the $B_t$ factor, where it can be
estimated as a small decrease of 3\% when $M_{\rm SUSY}$ is raised from
$167\,\GeV$ to $1\,\TeV$.

\medskip

Before closing this section, it is also interesting to discuss and
display the RG running of the determinants, from $M_Z$ to the GUT
scale. We remind that the determinants of the Yukawa matrices are not
RG invariant: for example for $\tb<25$ they follow the equations:
\be
\frac{\rm d}{{\rm d}t} \ln \det \hat \l_f=G_f^i\,\a_i+C_f\,\a_t\,,
\ee
where $t=4\pi\ln(\mu/\GeV)$, $\a_{i=1,2,3}$ are the gauge couplings
and $\a_t=\l_t^2/4\pi$.  The beta function coefficients
$C_{u,d,e}=\{12,1,0\}$ are due to ``top'' loops and
$G_u=\{-13/5,-9,-16\}$, $G_d=\{-7/5,-9,-16\}$, $G_e=\{-27/5,-9,0\}$
are due to gauge loops.\footnote{For the (non-supersymmetric) Standard
Model, $C_{u,d,e}=\{21/2,15/2,9\}$, while $G_u = \{-51/20, -27/4,
-24\}$, $G_d= \{-3/4, -27/4, -24\}$, $G_e= \{-27/4, -27/4, 0\}$.}

\begin{figure}[t]%
  \centerline{\epsfig{file=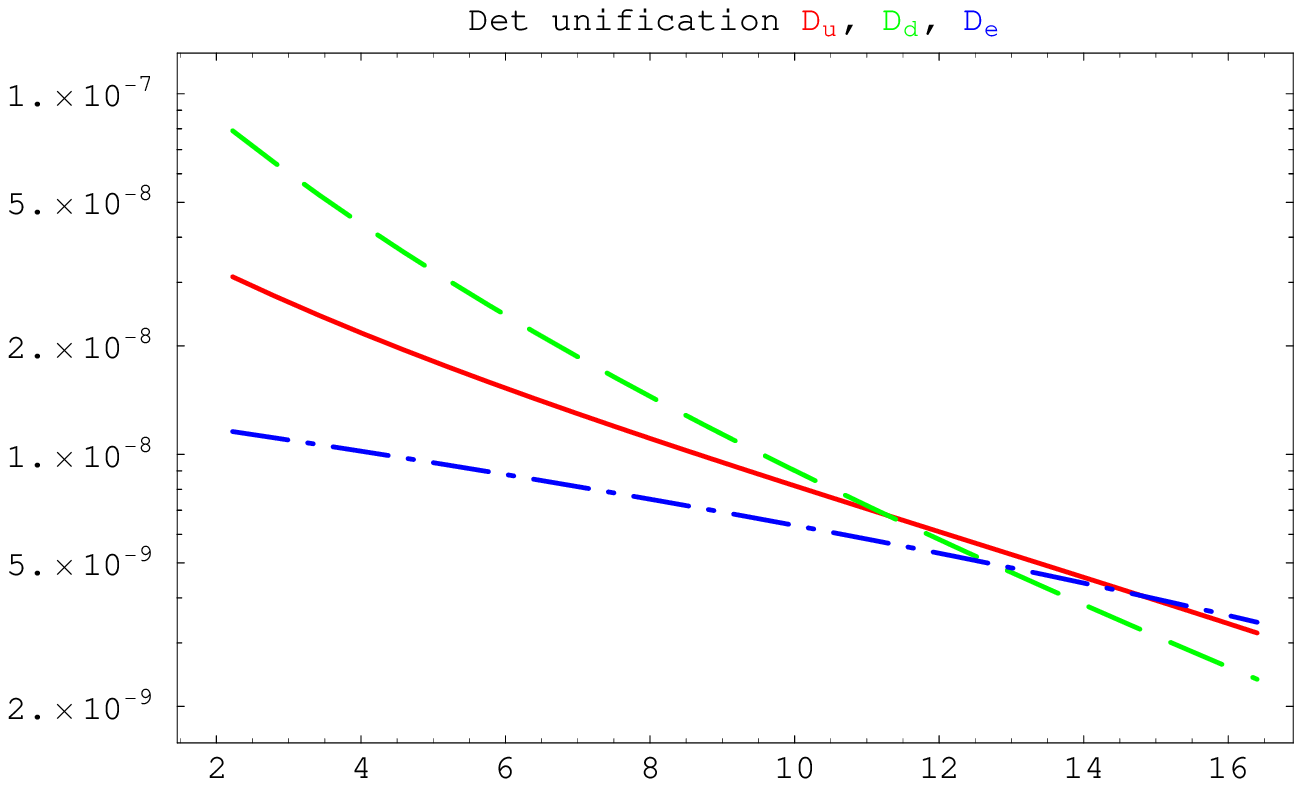,height=13em,width=19em}~~~~~
              \epsfig{file=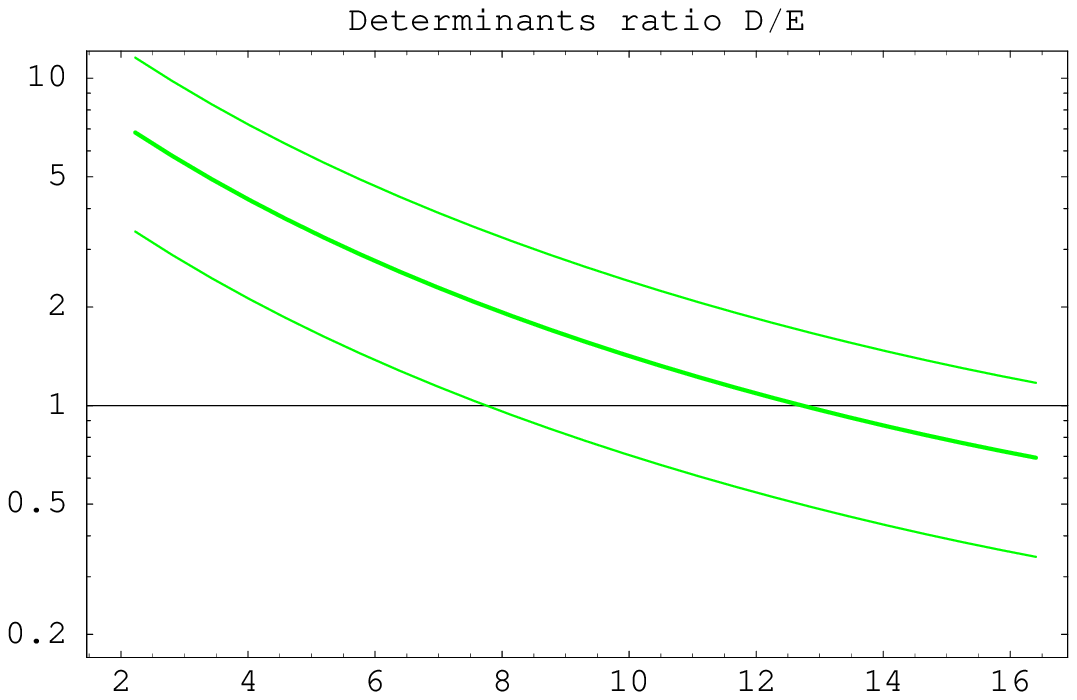,height=13em,width=19em}}
  \caption{Renormalization of Yukawa determinants, from $M_Z$ to
  $M_{GUT}$ in the MSSM, assuming $\tb\simeq8$. The left plot shows
  the three Yukawa determinants for U (red, continuous), D (green,
  dashed), E (blue, dot-dashed).  The right plot is more quantitative
  and shows the D/E ratio, with the one-sigma uncertainty.  It becomes
  compatible with 1 beyond $10^8\,\GeV$.\label{fig:detunif}}
\end{figure}

One can check analytically that these coefficients allow the
unification of the D, E determinants at high scale, within their
uncertainties.  For illustration we plot in figure~\ref{fig:detunif}
how the determinants are renormalized from $M_Z$ to the GUT scale, for
$\tb\simeq8.5$.  In the right plot we show the ratio of the D and E
determinants, as a test of the unification, given that the U sector is
used to define $\tb$ as in (\ref{eq:predtb1}), (\ref{eq:predtb2}). We
show also the one sigma uncertainties, defined when all ``down''
masses are deviated simultaneously at their one-sigma limit.  The
picture shows indeed an approximate unification of determinants at
high energy, since the curve is compatible with the value of 1 for
scales above $10^8$, to $10^{16}\,\GeV$, with a slight preference for
a scale of $\sim 10^{13}\,\GeV$. At the GUT scale the unification is
precise within $25\%$, that is less than the uncertainty on the D
determinant (see indeed (\ref{eq:predmdms})).

\medskip

We also comment on the results of a similar analysis in the case of
the non-supersymmetric Standard Model. We find an even better
agreement of the D, E determinants (within 2\%) at the scale of the
$\a_1$-$\a_2$ unification, $M_{\mbox{\scriptsize
GUT-12}}\simeq10^{13}\,\GeV$.  However, even ignoring the lack of
simple unification of the three gauge coupling constants, the
requirement of the U determinant to match with D, E requires an
extension to models of the 2HDM type, with the relative $\tb$
parameter again in the moderate range. It would be interesting to
check these ideas in such a non supersymmetric model that unifies via
an intermediate scale.

\section{Model realization: Rank-1 combinations in flavour space}

In this section we describe a model that realizes the above ideas in
the framework of SO(10) SUSY GUT with SU(3) flavour symmetry.  The
question to be answered is \emph{``what in this framework generates
fermion mass hierarchies and flavour mixing?''}. Indeed one can use
the fact that some hierarchic parameters are generated by the SO(10)
breaking: they distinguish between the U, D, E sectors, but
independently of the flavour. Then one can imagine that when flavour
is broken these hierarchic parameters will be assigned to the correct
direction, i.e.\ large ones in the third family direction in all
fermion sectors.  Since the unification of determinants observed in
the previous section seems to suggest that some relations among
eigenvalues may not depend on the fermion sector, and also do not
depend on the flavour mixing, we can then try to think that in first
approximation it is possible to disentangle the hierarchy of
eigenvalues from the mixing structure, that in turn can be common to
the U, D, E sectors.  Once a flavour breaking is introduced it will
not only generate mixings, but, if it is not just unitary matrices, it
will also modify slightly the eigenvalues, and thus will generate the
deviations $A_{e,d}$.

Following~\cite{BN} we review here how these ideas can be implemented,
writing the Yukawa matrices at GUT scale as a hierarchic combination
of three fixed rank-one projectors in flavour space. This approach was
devised in~\cite{br,z} as a description of an \emph{inverse hierarchy}
unification at very low $\tb$, as in the first case of
figure~\ref{fig:abc}.  It turns out however that it is quite
successful in accounting for fermion masses and mixings even at
moderate $\tb$, in fact it is even more successful.

Let us start with the case of no mixings and ignoring the deviations
from strict hierarchy.  As mentioned in the beginning, it is clear
that this case is quite simple, and one would build a model that
produces the following Yukawa matrices:
\bea
  \hat\l_f^{\rm {no\ mix}}\!\!\! &=&{}\l\,\left[\e_f\matr{1&0&0\\0&0&0\\0&0&0}\,+
    \matr{0&0&0\\0&1&0\\0&0&0}\,+\e_f^{-1}\matr{0&0&0\\0&0&0\\0&0&1}\right],\quad f=u,d,e
\0
\eea
with $\e_{e,d}\simeq1/50$ and $\e_u\simeq1/500$. Notice that the
determinant here is independent of the fermion sector $f$, i.e.\
$\det\l_f=\l^3$, while the eigenvalues follow different hierarchies in
different sectors. Unification of determinants follows from choosing
\{$\e_f$, 1, $\e_f^{-1}$\} coefficients.\footnote{We anticipate that
since these are built as effective operators, there are two other
allowed choices that shift the exponents by $\pm1$ power of $\e_f$,
leading to the other cases of figure~\ref{fig:abc}. For example
\{$\e_f^2$, $\e_f$, 1\} leads to the third case. The first and the
third case do not unify the determinants but may be analyzed
similarly. They however are in high tension with the experimental
data: they predict mainly low $m_t$; see~\cite{BN}.}  This form
ignores the deviations as well as the mixings, but the reason of this
is evident: three chosen matrices are mutually orthogonal projectors.

One is therefore led to consider a modification of the above, using
the most general rank-one projectors, in place of orthogonal ones.  It
is necessary that they are rank-one, because the eigenvalues should
reflect the hierarchy of the $\e_f$.

The Yukawa matrices are thus taken as a hierarchic combination of
three generic rank-one projectors in flavour space $\hat P_{1,2,3}$:
\bea
  \hat\l_f &=&{}\l\,\Big[\e_f\,\hat P_1\,+\hat
    P_2\,+\e_f^{-1}\,\hat P_3\Big]\,,\qquad f=u,d,e,\nu_D
\label{eq:dirgeneric}
\eea
where we have included also the Yukawa matrix for neutrinos, that will
anyway be present in SO(10) unification, and we will consider it
later.  The $\hat P_{n=1,2,3}$ are three generic rank-one complex
matrices and may be parametrized as $\hat P_n=p_n\otimes
p'_n=p_n{p'}_n^t$, with $p_n$, $p'_n$ three couples of vectors in
flavour space (flavour triplets).  The $p_n$ and $p'_n$ may be
realized as the VEV of {\em flavour triplets}, and may be thought as
providing the 'left' and 'right' flavour structures. They are thus
responsible for the breaking of the SU(3) flavour symmetry, and this
breaking does not depend on the fermion sector. Once both SO(10) and
SU(3) are broken, the $\e_f$ hierarchies are mixed via the $\hat
P_n$'s to generate both almost hierarchic eigenvalues as well as
mixing angles.

In this way one has a cooperation of the gauge (vertical) and flavour
(horizontal) sectors, of the theory, while allowing quite general mass
matrices.

Looking at the determinant of these $\hat\l_f$ matrices, it turns out
to factorize as:
\be
\det\hat\l_f \;=\;\l^3\,D_p\,D_{p'}\,,
\label{eq:det}
\ee
with $D_p=p_1\cdot(p_2\wedge p_3)$ and similarly for $D_{p'}$. This is
again independent of $\e_f$, i.e.\ the determinants are unified in the
U, D, E, $\nu$ sectors and the eigenvalues follow the predictions of
the previous sections, independently of the parameters entering in
$p_n$ and $p'_n$.

\medskip

To see in detail how this ansatz works, we choose a flavour basis and
we restrict for simplicity to the case of symmetric projectors for
example assuming that $p'_n=p_n$ are the VEV of the same triplet:
\bea
\label{eq:dir}
  \l_f = \l\Big[\e_f\hat P_1\,+\hat P_2\,+\e_f^{-1}\hat P_3\Big]&\!\!=\!\!&\l\left[
 \e_f  \matr{1&a&b\\a&a^2&ab\\b&ab&b^2}
+\!      \matr{0&0&0\\0&1&c\\0&c&c^2}
+\e_f^{-1} \matr{0&0&0\\0&0&0\\0&0&d^2}
                           \right]\0\\
&\!\!=\!\!&\l
\arraycolsep=.8em
\matr{\e_f& a\,\e_f&b \,\e_f \\[1ex]
a\,\e_f & 1+a^2\e_f& c+ab\,\e_f\\[1ex]
b\,\e_f & c+ab\,\e_f & d^2/\e_f+c^2+b^2\e_f}\,,
\eea
where we have rotated globally all the $p_n$ vectors in flavour space
and rescaled $\e_f$, $\l$ to set $p_1=(1,a,b)$, $p_2=(0,1,c)$,
$p_3=(0,0,d)$. Eliminating also the irrelevant phases, we end up with
the following generic 13 parameters domain: $\l,a,d\in {\mathbb R}^+$,
$ b,c,\e_u,\e_d,\e_e\in{\mathbb C}$. These parameters should account
for all Dirac couplings of charged fermions.

\medskip

It is also worth displaying an other form of this matrix
\newcommand\bmaz{\begin{array}{c|c|c}}
\newcommand\emaz{\end{array}}
\bea
  \hat\l_f &=&{}\l\,\hat
  T\matr{\e_f&0&0\\0&1&0\\0&0&\e_f^{-1}}\hat T^{\,t}\,,\qquad \hat
  T=\left(\bmaz &&\\p_1&p_2&p_3\\&& \emaz\right)=\matr{1&0&0\\a&1&0\\b&c&d},
\label{eq:TET}
\eea
where we can see that the flavour breaking triplets $p_n$ induce a
flavour mixing through $\hat T$ that is not unitary, and therefore
will induce not only mixing angles, but also deviations in the
eigenvalues. The deviations will therefore be linked to the flavour
mixing. This form is also useful to easily prove~(\ref{eq:det}).

\subsection{Leading order analysis}
\label{sec:firstorder}

The form (\ref{eq:dir}) can be taken as an ansatz and can be solved
quite completely by computing the leading order expressions of eigenvalues
and mixing angles:
 \bea
\label{eq:yukup}
 \l_u\simeq\l\,|\e_u|\,,\ \ \qquad\quad&\quad&\l_c\simeq\l\,,\quad\ \ \ \ \ \
 \qquad\qquad\l_t\simeq\l\left|\frac {d^2}{\e_u}\right|
\\[.7em]
 \l_d\simeq\l\frac{|\e_d|}{|1+{ a^2\e_d}|}\,,\quad&\quad&\l_s\simeq\l|1+{
 a^2\e_d}|\,,\quad\qquad\l_b\simeq\l\left|\frac {d^2}{\e_d}\right|
\\[.7em]
 \l_e\simeq\l\frac{|\e_e|}{|1+{
 a^2\e_e}|}\,,\quad&\quad&\l_\mu\simeq\l|1+{a^2\e_e}|\,,\quad\qquad\l_\tau\simeq\l\left|\frac {d^2}{\e_e}\right|
 \eea
 \vspace*{-.3em}
 \be
 \t_{12}\simeq-\frac{a\e_d}{1+a^2\e_d}\,,\qquad\t_{23}\simeq 
 \frac{\eta_d}{d^2}\e_d\,,\qquad \t_{13}\simeq \frac b{d^2} \e_d^2\,,\qquad
  \gamma_{CP}\simeq\arg\left(\frac{b (1+a^2\e_d)}{\eta_d}\right).
 \ee
where $\eta_d=(c+ab\, \e_d)$, and since $\e_u\ll\e_{d,e}$, the quark
mixing angles are well approximated by the ``down'' sector.  We will
not repeat the whole analysis that can be found in~\cite{BN} but just
quote the main steps that demonstrate how the horizontal and vertical
hierarchy cooperate.

\begin{itemize}
\item[-] First from $\l_b\simeq\l_\tau$, one should have
  $|\e_e|\simeq|\e_d|$ as already discussed.

\item[-] Second, one notes that, ignoring the corrections, the
  second generation Yukawa are unified $\l_{c,s,\mu}\simeq\l$.

\item[-] However, the RG invariant relation
  $s_{12}\simeq\sqrt{\Md/\Ms}$ tells us that $|a^2\e_d|\simeq1$, so
  that $a$ should actually be quite large, $|a|\sim|\e_d|^{-1/2}$
  ($\simeq7$, see later).  

  This large value of $a$ generates the right Cabibbo angle, but also
  enters the correction of the eigenvalues in the first two
  generations of D, E sectors.

\item[-] Indeed right splitting of s,$\mu$ Yukawa is produced by the
  factors $A_{d,e}=|1+a^2\e_{d,e}|$ once the sign (the phases) of
  $a^2\e_e$ and $a^2\e_d$ are nearly opposite,
  $a^2\e_e\simeq-a^2\e_d\simeq\mbox{e}^{0.5i}$, so that
  $|A_e|\simeq2$, $|A_d|\simeq0.5$. As discussed above, for the first
  generation the split goes (correctly) in the opposite direction,
  realizing the small flavour seesaw.

  So, in this ansatz, the largeness of $a$, following from the
  Cabibbo angle, explains also the deviation from exactly hierarchical
  family masses of $e$, $d$, $\mu$, $s$. This was already observed
  in~\cite{br}.

\item[-] From other known mass ratios one can then determine
  $|\e_{e,d}|\simeq1/55$, $a\simeq7.5$, $|\e_u|\simeq1/600$,
  $d^2\simeq0.5$, $\tb\simeq7.3$. Then further insight can be gained
  by checking the CP phase and the other mixing angles: from these it
  turns out that also $b$ is quite large, like $a$, i.e.\
  $|b|\simeq5$. Also, one finds two branches: in one the complex $b$
  and $c$ are almost orthogonal, in the other they are
  opposite. Finally, there is a flat direction, a combination of
  $|\e_e/\e_d|$ with the common phase of $b$, $c$.

\end{itemize}
Summing up, the ansatz is solved successfully and shows a one
parameter family of solutions, with two branches.  The precise
leading-order solution for all parameters is not conclusive since
next-to-leading corrections and contributions to CKM from the ``up''
sector are 10\% in magnitude, larger than experimental errors; hence a
numeric fit is necessary and will be described below.  Nevertheless
the ansatz is capable of accounting for all the charged fermion masses
and mixings, unifying the determinants and giving an additional
mechanism that links the deviations $A_{e,d}$ with the Cabibbo angle
$\t_{12}$.

\medskip

Before describing a model that realizes this ansatz, we would like to
comment on the solution found from the charged fermions so
far. Collecting the three $p_n$ vectors:
\bea
p_1=(1,\,7,\,5)\,,\qquad &p_2\simeq(0,\,1,\,0.2\mbox{--}1.5)\,,& \qquad p_3\simeq(0,\,0,\,0.7)\,,
\label{eq:vectors}
\eea
we observe that they tend to lie in the 2-3 plane.  Exact planarity is
broken to order $1/8\simeq\sqrt{\e_d}$, as can be seen from the first
1 with respect to the modulus of its vector $|p_1|\simeq 8.5$.  The
reason for this is to be tracked in the magnitude of the quark mixing
angles that require $a$ and $b$ to be large while $c$ and $d$ are of
order one.%
\footnote{We observe also that if one may be concerned with the moduli
of $p_n$ not being equal and of order 1, (they are in fact
$|p_1|\simeq 8$, $|p_2|\simeq1.5$, $|p_3|\simeq0.7$), they can be
brought to be in the range 1--2 by rescaling all the $\e_f$ by a
factor of 10. With this normalization the hierarchy of eigenvalues is
due partially to the $\e_f$ and partially to the $p_n$ vectors having
hierarchic entries, while the normalization that we adopted above
corresponds to an eigenvalue hierarchy that comes only from the
$\e_f$'s.  The choice may be varied depending on the theoretical
realization.}

We find this pattern of quasi-planar triplets a nice hint for the
realization of a flavour sector of the theory, where the $p_n$ vectors
are generated by the VEV of three scalar fields that are triplets of
the flavour SU(3).  The potential and the radiative corrections to it
lead to their mutual disorientation, along the liens
of~\cite{disoriented, sextets}. Incidentally, we also note that while
we started from complex triplets, fitting fermion masses and mixings
indicates that the only two complex parameters $b$, $c$ have a common
phase.  This means that the three triplets VEV can be taken completely
real. The CP violation is then produced just by the complex $\e_f$ and
the coupling constants.  This is most welcome, since in usual VEV
configurations, the triplets tend to repel or attract each other and
it is hard to obtain complex relative angles.  We leave this analysis
and the realization of the flavour sector for future work.

\medskip



In the next section we complete the description of the model in the
SO(10) side, where the rank-one projectors are mixed with the SO(10)
breaking VEVs that generate the hierarchy parameters.

\section{SO(10) universal seesaw and flavour}

Since the different $\e_{u,d,e}$ hierarchies correspond to a broken
SO(10), it is natural to generate them at SO(10) breaking, i.e.\ at
the GUT scale, and then transfer them in some way to the light
fermions-higgs Yukawa couplings.  At the GUT scale we should also mix
them with the flavour triplets VEVs $p_n$ mentioned above. We describe
here a way to implement this program, along the lines of~\cite{z,BN}.

\subsection{Small parameters from SO(10) breaking}

First we recall that among the different schemes of SO(10) breaking, a
very successful and economical one is that implementing the so called
missing-VEV mechanism~\cite{dimwilc}, that while breaking to the SM
gauge group, automatically stabilizes the hierarchy of the light
doublets with respect to the GUT scale. Among other fields, in this
mechanism there are three multiplets in the $\R{45}$ representation:
$\R{45}_X$, $\R{45}_R$, $\R{45}_{BL}$. Altogether they break SO(10) to
the SM group, therefore their VEVs are, in Pati-Salam notation, only
allowed along the two directions $(\R{15},\R1,\R1)$ and
$(\R1,\R1,\R3)$. It turns out that
$\langle\R{45}_R\rangle\in(\R1,\R1,\R3)$,
$\langle\R{45}_{BL}\rangle\in(\R{15},\R1,\R1)$ and
$\langle\R{45}_X\rangle\in(\R{15},\R1,\R1)+(\R1,\R1,\R3)$.

All these VEVs can be used in a ratio to some higher scale (e.g.\
$M\sim10^{17}\GeV$) to introduce small parameters into the game.
However the last one, the VEV of $\R{45}_X$, is of particular interest
since it depends generically on two coefficients, that define its
mixture of $(\R{15},\R1,\R1)$ and $(\R1,\R1,\R3)$. When the small
ratio $\langle\R{45}_X\rangle/M$ is projected onto the fermion
multiplets and it decomposes in four small parameters $\e_u$ $\e_d$
$\e_e$ $\e_{\nu_D}$, we have that only two of these are independent,
and they turn out~\cite{z} to be constrained by the following
relations:
\be
\e_e=-\e_d-2\e_u\,,\qquad\e_{\nu_D}=-2\e_d-\e_u\,.
\label{eq:soten}
\ee 
The first equation tells us that once $|\e_u| \ll|\e_d|$, the equality
in the D,E sectors $|\e_e|\simeq|\e_d|$ is predicted, nicely
reproducing the pattern observed in nature. Moreover, the same
relation implies also bottom-tau unification, with a deviation of
$\e_u/\e_e\simeq 10\%$, that is indeed observed.

The second equation gives a prediction on the Yukawa hierarchy in the
neutrino sector, and implies that it is milder than that of the
leptons, $\e_{\nu_D}\simeq1/25$.

Therefore without introducing new fields in the theory, we found that
the $\R{45}_X$ generates the right pattern of small hierarchy
parameters $\e_u$, $\e_d$, $\e_e$, $\e_{\nu_D}$.  For this reason we
will assume below that the flavour hierarchies are generated as
effective operators by powers of $\langle\R{45}_X\rangle/M$.

\subsection{SO(10) universal seesaw}

To generate the Yukawa couplings as the correct effective operators,
it is necessary to introduce new (super)heavy fermion multiplets, that
act as messengers. Let us start with three families of heavy
vectorlike fermions $\R{16}'_i$, $\Rb{16}'_i$.  To generate the light
Yukawa couplings they should couple both to the light fermions and to
the Higgs field, that we take to be the usual
$\R{10}=T(\R6,\R1,\R1)+\phi(\R1,\R2,\R2)$. This contains the MSSM
higgses in $\phi$.  The three families of light fermions then receive
an effective Yukawa coupling matrix from the mass matrix of heavy
vectorlike fermions $\R{16}'_i$, $\Rb{16}'_i$, via a ``universal''
seesaw mechanism~\cite{universal}. In detail, the superpotential contains
\be
\label{eq:pot}
W=\gamma\,\R{10}\, \R{16}^{i}\R{16}'_i \,+ \gamma'
\,\R{45}_R\,\R{16}^{i}\,\Rb{16}'_i\,+\Rb{16}'_i\,M_{heavy}^{ij}\R{16}'_j
\ee
where $\R{10}$ is the fundamental Higgs and $\R{45}_R$ is an adjoint
one whose role will be clear in a moment. Notice that by SU(3) flavour
symmetry a direct Yukawa coupling $\R{16}\R{16}\R{10}$ is
forbidden. The flavour structure is encoded in $\hat M_{heavy}$, while
the first two terms are flavour universal.  Near at the GUT scale, the
higgses develop a VEV, so that the heavy fermions $\R{16}'$ decouple
and the light fermions $\R{16}$ acquire an effective Yukawa coupling
with $\R{10}$ that is approximately given by the ``seesaw'' formula:
\be
\label{eq:allseesaw}
\hat \l=\gamma \,\gamma' \, \langle\R{45}_R\rangle\,\hat M_{heavy}^{-1}\,,
\ee
Therefore the Yukawa couplings of the light fermions $\l$ are 
proportional to the \emph{inverse} of the heavy fermion masses, that 
in turn should exhibit an \emph{inverted hierarchy} pattern, with the 
lightest particle being the heavy correspondent of the ``top''. 

\medskip


To make explicit the universal seesaw following from~(\ref{eq:pot}),
we decompose in Pati-Salam notation the fermion fields as:
$\R{16}^i=f^i(\R4,\R2,\R1)+f^{ic}(\Rb4,\R1,\R2)$ and
$\R{16}'_i={\cal F}_i(\R4,\R2,\R1)+F^c_i(\Rb4,\R1,\R2)$,
$\Rb{16}'_i={\cal
F}^c_i(\Rb4,\R2,\R1)+F_i(\R4,\R1,\R2)$. Note that $f=q,l$ and
${\cal F}=q',l'$ are isospin doublets and $f^c=u^c,d^c,e^c,\nu^c$,
$F=u',d',e',\nu'$, are isospin singlets. With this decomposition the
couplings can be illustrated as follows:
$$
\def\arraystretch{1.4}
\begin{array}{ccc}
 & {\begin{array}{ccc} \,f^c & \,\,\,\;F^c & \,\,\;{\cal F}^c
\end{array}}\\ \vspace{2mm}
\begin{array}{c}
f \\ F \\ {\cal F}   \end{array}\!\!\!\!\!&{\left(\begin{array}{ccc}
0 & {\gamma}\phi  & 0 \\ {M}_R  & \hat{M}_F & 0 \\
{\gamma}\phi
 & 0 & \hat{M}_{\cal F} \end{array}\right)}
\end{array}
\vspace*{-1ex}
$$
where $M_R$ is the scale of $\langle\R{45}_R\rangle$ and $\hat
M_{heavy}$ is decomposed in $\hat M_F$ and $\hat M_{\cal F}$ ,
respectively the mass matrices of heavy isosinglets and isodoublets
(unmixed).

Only the $F$, $F^c$ participate in the decoupling process and in each
sector the Yukawa matrix is proportional to the inverse of $\hat M_F$
alone:
\be
\label{eq:Fseesaw}
\hat\l_{f}=\gamma \,M_R \,\hat M_F^{-1}\,,
\rlap{$\qquad\qquad f=u, d, e, \nu_D\,.$}
\ee

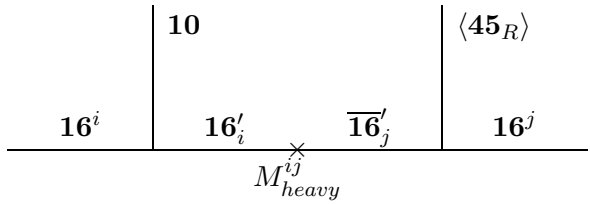
\begin{figure}[t]%
\centerline{
    \setlength\unitlength{5em}
    \newcommand\clap[1]{\hbox to 0pt{\hss#1\hss}}
    \newcommand\cclap[1]{\vbox to 0pt{\vss\hbox to 0pt{\hss#1\hss}\vss}}
  \begin{picture}(4,1.6)(.1,-0.4)
    \put(.5,.1){\clap{$\R{16}^i$}}
    \put(0,0){\line(1,0){1}}
    \put(1,0){\line(0,1){1}}
    \put(1.1,.8){$\R{10}$}
    \put(1.5,.1){\clap{$\R{16}'_i$}}
    \put(1,0){\line(1,0){1}}
    \put(2,0){\cclap{$\times$}}
    \put(2,-.2){\cclap{$M^{ij}_{heavy}$}}
    \put(2,0){\line(1,0){1}}
    \put(2.5,.1){\clap{$\Rb{16}'_j$}}
    \put(3,0){\line(0,1){1}}
    \put(3.1,.8){$\langle\R{45}_R\rangle$}
    \put(3,0){\line(1,0){1}}    
    \put(3.5,.1){\clap{$\R{16}^j$}}
  \end{picture}
}
\vspace*{-1em}
  \caption{Universal seesaw mechanism using $\R{45}_R$.\label{fig:seesaw}}
\end{figure}

It is important to note that the $f$-${\cal F}^c$ entry is zero
because $\langle\R{45}_R\rangle$ is in the ``right'' direction.  This
shows three remarkable features of this universal seesaw using
$\R{45}_R\in(\R1,\R1,\R3)$: 1) only the isosinglets $F^c$ participate
in the seesaw mixing with $f^c$, so that their mass matrix is directly
reproduced in the light Yukawa; this avoids a mixing with $M_{\cal F}$
that could spoil the exact hierarchies; 2) the LLLL (dominant) part of
the D=5 proton decay is automatically eliminated, because
(\R1,\R1,\R3) does not couple the light and heavy
isodoublets~\cite{z,proton}; 3) finally, again because the heavy
isodoublets do not enter the seesaw, the squark and sleptons mass
matrices are automatically aligned with the (square of the) Yukawa
matrices, and this eliminates problems of flavour changing neutral
currents.  We conclude that here the universal seesaw via the
SU(2)$_R$ breaking automatically allows the correct mass generation,
suppresses proton decay and SUSY flavour problems.

\subsection{Entangling flavour and SO(10) higgses}

We now come to the generation of the flavour structure in $\hat
M_{heavy}$.  We have already introduced the three flavour projectors
$P_n$ that should be coupled to different powers of the $\R{45}_X$
field.  Since however from~(\ref{eq:allseesaw}),~(\ref{eq:Fseesaw})
the heavy mass matrix is the inverse of the Yukawa couplings, we must
take advantage of a property of combinations of rank-one projectors:
the inverse of such a combination is a combination of new projectors,
with inverse coefficients:
\be\label{eq:inversion}
     \big(\a_1\hat P_1+\a_2\hat P_2+\a_3\hat
     P_3\big)^{-1}=\left(\frac1{\a_1}\hat Q_1+\frac1{\a_2}\hat
     Q_2+\frac1{\a_3}\hat Q_3\right),
\ee
where the $\hat Q_n$ are the ``reciprocal'' of the $\hat P_n$
projectors: if $\hat P_n$ are parametrized with flavour triplets as
$\hat P_n=p_np_n^t$, then  $\hat Q_n=q_nq_n^t$ are given by three
new triplets $q_n$ that are in fact the reciprocal of the $p_n$ ones:
$q_n=\frac12\e_{nmr}p_m\wedge p_r/D_p$, with $D_p=p_1\cdot(p_2\wedge
p_3)$.

Therefore one should generate a $\hat M_{heavy}$ as a combination of
three rank-one projectors $Q_n$ multiplied by powers of
$\R{45}_X/M$. To have hierarchic eigenvalues these should be
successive powers, i.e.:
$$
\hat M_{heavy} \propto 
\Big[\hat Q_1 +\frac{\R{45}_X}{M}\hat Q_2
    +\frac{\R{45}_X^{\!2}}{M^2}\hat Q_3\Big]\,,
$$
We do not specify at this point the effective realization of this
operator, that will be addressed later on. We only observe that such
effective operators may involve integer powers (positive or negative)
of fields, and therefore one may have in front a generic power of it:
\be
\label{eq:Mgen}
\hat M_{heavy} \propto 
\Big(\frac {\R{45}_X}M\Big)^{\!-n} 
\Big[\hat Q_1 +\frac{\R{45}_X}{M}\hat Q_2
    +\frac{\R{45}_X^{\!2}}{M^2}\hat Q_3\Big]\,,
\ee
It is clear that when at SO(10) breaking $\langle\R{45}_X\rangle/M$
generates the small parameters $\e_u$, $\e_d$, $\e_e$,
$\e_{\nu_D}$, one will receive effective Yukawa couplings in the form:
\be
\label{eq:Lgen}
  \hat\l_f=\frac{\e_f^{n-2}}{\l}\big[\e_f^2P_1+\e_fP_2+P_3\big]\qquad\qquad
  f=u,d,e,\nu_D
\ee

The interesting cases are $n=0$,$1$,$2$, that correspond to the three
cases of figure~\ref{fig:abc}:
\begin{itemize}  
\item[a)] with $n=0$, the coefficients are $1,\e_f^{-1},\e_f^{-2}$, and the
GUT Yukawa couplings of the \emph{first} generation are approximately
unified. This requires very low $\tb$ ($<2$);
\item[b)] with $n=1$, the coefficients are\ $\e_f,1,\e_f^{-1}$, we have an approximate
  unification of the \emph{second} generation, at intermediate
  $\tb$ ($\simeq10$);
\item[c)] with $n=2$, the coefficients are\ $\e_f^{2},\e_f^,1$, and large
$\tb$ ($>50$), one reaches unification of the \emph{third} generation,
as in the known case of t-b-$\tau$ unification.
\end{itemize}
The second case is the most interesting since there is optimal
agreement with the data, and as discussed in the present work, the
determinants are unified.  For the first case see~\cite{z}, for a
unified treatment see~\cite{BN}.

\medskip

Sticking to the interesting case of n=1 (b), we end this section by
providing the complete realization with renormalizable operators. Such
a realization is not unique, so we display two different ways to
achieve the same result. 

Every realization of nonrenormalizable operators like~(\ref{eq:Mgen})
requires one to introduce additional fields. For example one can
introduce four more vector-like fermion multiplets that are
\emph{flavour singlets}: $\R{16}''_1$, $\R{16}''_2$, $\R{16}''_3$ and
$\R{16}'''$ (and their conjugates); as an other example one can use in
place of three triplets $q_n$, two triplets $q_{1,2}$ plus one sextet
$\chi_3$, and then introduce three vector-like fermion fields: two
flavour singlets $\R{16}''_{1,2}$ and one triplet
$\R{16}^{\prime\prime\prime i}$ (and their SO(10)$\times$SU(3)
conjugates). The coupling are arranged as follows in these two cases:
$$
\def\arraystretch{1.3}
 \ba[t]{l|cccccc|}
 &\R{16}^i&\R{16}'_i&\R{16}''_1&\R{16}''_2&\R{16}''_3&\R{16}'''\\
 \hline
 \R{16}^i &   &\R{10}&&&&\\
 \Rb{16}'_j &\R{45}_R&   &q_1^j&q_2^j&q_3^j&0\\
 \Rb{16}''_1&   &q_1^i&\R{45}_X&&&\\
 \Rb{16}''_2&   &q_2^i& &{M}&&\\
 \Rb{16}''_3&   &q_3^i& &&&{M}\\
 \Rb{16}'''& &0&&&{M}&\R{45}_X\\
 \hline
 \ea
\qquad
\def\arraystretch{1.3}
 \ba[t]{l|ccccc|}
 &\R{16}^i&\R{16}'_i&\R{16}''_1&\R{16}''_2&\R{16}^{\prime\prime\prime\,i}\\
 \hline
 \R{16}^j &   &\R{10}&&&\\
 \Rb{16}'_j &\R{45}_R&   &q_1^j&q_2^j&\R{45}_X\\
 \Rb{16}''_1&   &q_1^i&\R{45}_X&&\\
 \Rb{16}''_2&   &q_2^i& &{M}&\\
 \Rb{16}^{\prime\prime\prime}_j&   &\chi_3^{ij}& &&{M}\\
 \hline
 \ea
$$ 
In the first example $q_n$ define exactly the three reciprocal
triplets mentioned above, while in the second the $\hat Q_{1,2,3}$
projectors will be given by $\hat Q_1\propto q_1q_1^t$, $\hat
Q_2\propto q_2q_2^t$ and~$\hat Q_3\propto \chi_3$.

\medskip

One final remark is in order: in these examples some of the couplings
are forbidden. We do not address here the way to motivate this, that
can be done by means of additional symmetries, discrete or continuous.
Indeed the way that heavy higgses and flavour triplets/sextets are
coupled in these examples (and e.g.\ in~(\ref{eq:Mgen})) points
towards a further symmetry that explains the charges or couplings much
in the Froggatt-Nielsen spirit.

\section{Neutrinos}

In this section we analyze how the neutrino sector can be accommodated
in the present framework of flavour triplets coupled with SO(10) heavy
Higgs fields.  From the mechanism described above, the Dirac neutrino
masses are unified together with the other fermions and thus result in
the range 10--1000\MeV. Therefore we need to introduce also a Majorana
mass for right-handed neutrinos, and generate the light neutrino mass
by canonical seesaw~\cite{seesaw}.  Then given that in the model the
flavour structure is encoded in the three ${P}_n$ projectors, we
assume that also the Majorana masses are built with the same
projectors.

The Majorana mass for the RH neutrinos that belong to the light
$\R{16}$ multiplet is generated via an universal seesaw similar to
that of Yukawa couplings, this time via a singlet messenger. We do not
dwell in the details that may be found in~\cite{BN}, and we quote the
result, that also the RH neutrino mass $\hat M_{\nu_R}$ will be built
with the same flavour projectors $P_n$ and with a new hierarchic
parameter $\e_{\nu_R}$.  The coefficients however are generic, since
the three new coupling constants can no more be absorbed in the
$P_n$. We choose to make explicit the $\e_{\nu_R}$ and add a further
complex parameter $\a$.
\be
\label{eq:RH}
  \hat \l_{\nu_R}\propto\Big[ \P_1+\e_{\nu_R} \P_2+\frac{\e_{\nu_R}^2}\a \P_3\Big]\,,
\ee
where for simplicity here and in the following we prefer to use the
dimensionless matrices $\hat\l_{\nu_R,\nu_L}=\hat
M_{\nu_R,\nu_L}/v\sin\beta$, with the electroweak $v=174\,\GeV$.

A mass matrix for the left-handed neutrinos is then generated by
canonical seesaw as $\hat
\l_\nu=\hat\l_{\nu_D}\hat\l_{\nu_R}^{-1}\hat\l_{\nu_D}$, between the
neutrino yukawa~(\ref{eq:dir}) and the RH mass matrix~(\ref{eq:RH}).
The result can be found by exploiting an other interesting property of
rank-one combinations: indeed, the following ``seesaw'' relation
holds:
$$
\big(\a_{1}\, \hat P_1+\a_{2}\, \hat P_2+\a_{3}\,\hat P_3\big)\big(\b_{1}\, \hat P_1+\b_{2}\,
  \hat P_2+\b_{3}\,\hat P_3\big)^{-1}\big(\a_{1}\, \hat P_1+\a_{2}\, \hat P_2+\a_{3}\,\hat P_3\big)=
$$
\vspace*{-1em}
\be
=\left(\frac{\a_{1}^2}{\b_1}\, \hat P_1+\frac{\a_{2}^2}{\b_2}\, \hat P_2+\frac{\a_{3}^2}{\b_3}\,\hat P_3\right),
\label{eq:seesaw}
\ee 
i.e.\ the seesaw acts on the coefficients only, but does not change
the flavour projectors $\hat P_n$.  Hence, we get that
$\hat\l_{\nu_L}$ is built again as a combination of the same flavour
projectors:
\be
\hat\l_{\nu_L}=\hat\l_{\nu_D}\hat\l_{\nu_R}^{-1}\hat\l_{\nu_D}\propto \Big[\e_\nu^2\hat P_1+\e_\nu \hat P_2+\a \hat P_3\Big]\,,\qquad
\e_\nu=\frac{\e_{\nu_D}^2}{\e_{\nu_R}}\,.
\ee

Using the explicit form~(\ref{eq:dir}), we can rewrite:
%
$$
\l_\nu\propto
\matr{|\e_\nu/\bar A_\nu|&|a\e_\nu/A_\nu|&|b\e_\nu/A_\nu|\\
      |a\e_\nu/A_\nu|    &  1        &\rlap{$\rho$}\mbox{\phantom{$\rho^2 +\sigma$}}\\
      |b\e_\nu/A_\nu| & \rho      & \rho^2 +\sigma}.
$$
with $A_\nu=(1+a^2\e_\nu)$,~~$\rho= (c+ab\e_\nu)/A_\nu$,~~$\sigma=(d^2\a+(b-ac)^2\e_\nu^2)/A_\nu$.

In this matrix, we only have two free complex parameters $\e_\nu$ and
$\a$,\footnote{The common phase of $b$, $c$, that was a flat direction
of the charged fermions can be reabsorbed in the phase of $\a$.}
therefore the model is quite constrained since it must explain the
neutrino mass hierarchy and the three mixing angles.

Unlike the case of charged fermions, an exact analytical solution of
this matrix is not possible, due to the large neutrino mixings and to
the small but important uncertainties in the parameters determined
from the charged fermions.  Nevertheless it is possible to see that
neutrinos are predicted to be hierarchic, non degenerate, and to
derive a prediction for the $\t_{13}$ mixing angle. The analysis goes
as follows:
\begin{itemize}
\item[-] Non-degenerate neutrinos with $m_{\nu2}/m_{\nu3}\simeq1/5$
  follow from the fact that $a$ is large, so the ratio of the 1-1 with
  the 1-2, 1-3 entries is small $\simeq1/7$.

\item[-] Non-zero $\t_{e3}$ follows from $a>b$, so that the 1-2 entry
  is larger than the 1-3 one. An analytical estimate of the angles
  from the 2,3 sector can be given first in the form of a relation
  between them, using the parameters found from the charged sector and
  after imposing maximal $\t_{atm}$:
\be
\frac{\t_{e3}}{\t_{sol}} = \frac{m_{\nu2}}{m_{\nu3}}\,
\Big|\frac{1+R\,{\rm e}^{i\delta}}{1-R\,{\rm e}^{i\delta}}\Big|\,,
\qquad
\mbox{with }\ R=\frac{b}{a}\simeq\frac{\t_{13}^{quark}\l_b}{\t_{12}^{quark}\l_s}
\simeq \frac1{1.6 \thicksim 2.1}\,,
\ee
where $\delta=\arg \rho =\arg[(c+ab\e_\nu)/(1+a^2\e_\nu)]$. After imposing the
correct neutrino hierarchy and solving the model for $\e_\nu$, we can
estimate also:
\be
\t_{12}\simeq 30^\circ \Big|1-R\,{\rm e}^{i\delta}\Big|\,,\qquad 
\t_{13}\simeq 6^\circ \Big|1+R\,{\rm e}^{i\delta}\Big|\,.
\ee 
Hence, choosing  ${\rm e}^{i\delta}$ purely imaginary or just
slightly negative, we see that the right $\t_{sol}\simeq32^\circ$ is 
accomodated, with a prediction for $\t_{e3}\simeq4^\circ$.

The precise numbers are sensible to the quark mixing angles and
CP-phase, as well as to the neutrino hierarchy. In practice $\t_{e3}$
can be lowered to reach also $\t_{e3}\simeq0.5^\circ$ by stretching
within 1$\sigma$ the angles and the neutrino hierarchy.

\item[-] Finally, a new flat direction emerges, along which the angles
and hierarchy do not vary. It directly corresponds to the leptonic CP
violation phase, that may be predicted once angles and hierarchy will
be known to better accuracy. Also a majorana phase varies along this
flat direction.
\end{itemize}

Summing up, the framework naturally predicts hierarchic neutrinos with
nonzero $\t_{e3}$ in the 1$\sigma$ range, and there is a useful
conspiracy from the charged sector to allow the right neutrino
pattern.  A new flat direction emerges as a combination of the four
real parameters, to be added to the one found in the charged leptons
sector.  Therefore the model effectively takes advantage of just 13 of
the 15 real parameters to reproduce the known data.

\subsection{Numeric Fit and Results}

As mentioned above, we performed in~\cite{BN} a numerical analysis,
taking into account the exact renormalization factors. We performed a
best fit of neutrinos together with charged fermions. The $\t_{e3}$
neutrino mixing, that we have shown to be predicted inside its allowed
$1\sigma$ range, is not used as a fit condition, but rather displayed
as a model prediction. The fit was performed using the 15 parameters
(11 charged fermions, 4 neutrino) against all known data (20 tests).

\def\topfraction{1}
\def\bottomfraction{1}
\def\textfraction{0}

\begin{table}[t]
{\footnotesize
$$
\def\arraystretch{1.2}
\arraycolsep=.9ex
\begin{array}[t]{|c|c|c|l|}
\hline
& \mbox{exp} & \mbox{fit} & \multicolumn{1}{|c|}{\mbox{pull}} \\
\hline
m_t             & 167\pm5           &   169        & +0.4        \\
m_b             & 4.25\pm0.15       &   4.4        & +0.9        \\
m_c             & 1.25\pm0.1        &   1.3        & +0.4        \\
m_d             & 6\pm2             &   6.3        & +0.1        \\
m_s             & 105\pm25          & {\red 134}   & {\red +1.1} \\
m_s/m_d         & 19.5\pm2.5        &  21.4        &  +0.6    \\
\mbox{Ellipse}  & 23\pm2            & {     22.1 } & {     -0.4} \\
m_u/m_d         & 0.5\pm0.2         & {\red 0.25 } & {\red -1.2} \\
m_u\!\!+\!m_d   & 8.5\pm2.5         &   8.         & -0.2        \\
\hline
m_e             & 0.511             & .510         & <0.1        \\
m_\tau          & 1777              & 1779         & <0.1        \\
\hline
\end{array}
\qquad
\begin{array}[t]{|c|c|c|l|}
\hline
& \mbox{exp} & \mbox{fit} & \multicolumn{1}{|c|}{\mbox{pull}}\\
\hline
\theta_{c}      & 12.7\pm0.1        & 12.75       & +0.5        \\
V_{cb} 10^2\!\! & 4.13\pm0.15       & 4.14        & <0.1        \\
V_{ub} 10^3\!\! & 3.7\pm0.47        & 3.77        & +0.2        \\
\sin\!2\b_{CP}  & 0.736\pm0.05      & 0.734       & -0.1        \\
\gamma_{CP}     & 1.03\pm0.22       & 1.07        & +0.2        \\
\hline
\theta_{sol}    & 32\pm3            & 32.03       & <0.1        \\
\theta_{atm}    & 45\pm7            & 44.1        & <0.1        \\
h_\nu           & 0.18\pm0.07       & 0.185       & <0.1        \\
\hline
\hline
\blue \theta_{e3}     & \blue 0^\circ\mbox{ -- }9.6^\circ  & \blue 0.5\mbox{--}7.0& \blue (br1)   \\[-.2ex]
                &                                          & \blue1.7\mbox{--}4.3 & \blue (br2)   \\
\blue\tb             &                   & \blue 7.8         &             \\
\hline
\end{array}
$$
}
\caption{Best fit with $M_{SUSY}=m_t$. Total $\blue
  \chi^2=0.35$ per d.o.f.\label{tab:fit1}}
\end{table}

\begin{table}[t]
{\footnotesize
$$
\def\arraystretch{1.2}
\arraycolsep=.9ex
\begin{array}[t]{|c|c|c|l|}
\hline
& \mbox{exp} & \mbox{fit} & \multicolumn{1}{|c|}{\mbox{pull}} \\
\hline
m_t             & 167\pm5           &   168        & <0.1        \\
m_b             & 4.25\pm0.15       &   4.4        & +0.4        \\
m_c             & 1.25\pm0.1        &   1.3        & +0.4        \\
m_d             & 6\pm2             &   6.1        & +0.4        \\
m_s             & 105\pm25          &   127        & +0.8        \\
m_s/m_d         & 19.5\pm2.5        &   21.0       & +0.6       \\
\mbox{Ellipse}  & 23\pm2            & {     22.2 } & {     -0.4} \\
m_u/m_d         & 0.5\pm0.2         & {     0.32 } & {     -0.8} \\
m_u\!\!+\!m_d   & 8.5\pm2.5         &   8.         & -0.2        \\
\hline
m_e             & 0.511             & .510         & <0.1        \\
m_\tau          & 1777              & 1779         & <0.1        \\
\hline
\end{array}
\qquad 
\begin{array}[t]{|c|c|c|l|}
\hline
& \mbox{exp} & \mbox{fit} & \multicolumn{1}{|c|}{\mbox{pull}}\\
\hline
\theta_{c}      & 12.7\pm0.1        & 12.7        & <0.1        \\
V_{cb} 10^2\!\! & 4\pm0.15          & 4.13        & <0.1        \\
V_{ub} 10^3\!\! & 3.7\pm0.47        & 3.76        & +0.2        \\
\sin\!2\b_{CP}  & 0.736\pm0.05      & 0.734       & <0.1        \\
\gamma_{CP}     & 1.03\pm0.22       & 1.06        & +0.1        \\
\hline				                         	    
\theta_{sol}    & 32\pm3            & 32.0        & <0.1        \\
\theta_{atm}    & 45\pm7            & 45.0        & <0.1        \\
h_\nu           & 0.18\pm0.07       & 0.185       & <0.1        \\
\hline
\hline
\blue \theta_{e3}     & \blue 0^\circ\mbox{ -- }9.6^\circ  & \blue 0.5\mbox{--}7.0& \blue (br1)   \\[-.2ex]  
                &                                          & \blue1.6\mbox{--}4.2 & \blue (br2)   \\ 
\blue\tb             &                   & \blue 7.4         &             \\
\hline
\end{array}
$$
}
\caption{Best fit with $M_{SUSY}=1\,\TeV$ total $\blue\chi^2=0.14$ per
   d.o.f.\label{tab:fit2}}
\end{table}

The best fit results are shown in tables~\ref{tab:fit1}
and~\ref{tab:fit2}: the overall $\chi^2=0.35$ \emph{per d.o.f.}  is
quite good, and the model can account for all experimental
constraints, albeit having a bit low $m_u$ and a bit large $m_s$
(1$\sigma$ level) as implied by the predictions described in
section~\ref{sec:detunif}.

\smallskip

The fit results confirm the analytical study:
\begin{itemize}
\item[-] Apart from $m_u$, $m_s$ that deviate at most 1-$\sigma$, all
         data are fitted without tension ($<0.3\sigma$).
\item[-] The flat direction and the two branches correctly emerge from
         the fit. In the second branch all complex phases are almost
         aligned, hinting for a model with complex aligned $\e_f$'s
         and real triplets, with a reduction of the total number of
         real parameters to 10.

\item[-] Neutrinos are predicted as hierarchic and non degenerate, and
         the mixing from terrestrial neutrinos is nonzero:
         $\t_{e3}\simeq2.5^\circ,4.9^\circ$.

\item[-] Neutrino come out hierarchic and non degenerate
         $m_{\nu}\simeq(0.0045,\, 0.01,\, 0.048)\,\eV$, and
         right-handed neutrino masses are of the order ($10^7$,
         $10^9$, $10^{11}$) \GeV.

\end{itemize}


\section{Discussion}

In this work we discussed the structure of the Yukawa matrices as
renormalized at a common high scale, and in the context of MSSM GUT,
noticed the \emph{unification of their determinants}.  This new
unification translates in precise predictions in terms of charged
fermion masses: in particular it gives a prediction for $m_dm_s$ in
1$\sigma$ accordance with the experimental ranges, and a prediction of
$\tb\simeq7$ that lies in the currently favoured range.  In addition
this mechanism introduces a constraint that may explain why the first
family fermions are light when the third ones are heavy, in a kind of
flavour seesaw.

In fact, as opposed to traditional unification schemes, this idea does
not unify some combination of yukawa matrices at GUT scale, but
unifies approximately the yukawa eigenvalues of the \emph{second
family}.

\smallskip

Examining the mass matrices in more detail we presented an ansatz,
similar to the ones put forward in in~\cite{br, z}, that realizes the
unification of determinants by building the yukawa matrices as
hierarchical combinations of three generic rank-one projectors in
flavour space. This construction is successful in explaining the
deviations from exact mass hierarchy in the D, E sectors, linking them
to the Cabibbo mixing.

\smallskip

We reviewed how these ideas can be realized in a predictive model of
SU(3) flavour symmetry in the context of SO(10) SUSY GUT, along the
lines of~\cite{BN}.  In this model the mass matrices can be analyzed
with a leading-order analytical approach and with a complete numerical
fit. All fermion masses and mixings are accomodated at 1$\sigma$
level, including neutrinos, for which the analysis and the numerical
fit suggest direct and hierarchical neutrinos with non zero
$\t_{e3}=1.7^\circ$--$4.3^\circ$ or $0.5^\circ$--$7^\circ$, in two
branches.

The model uses the tools that are already available in the context of
SO(10) SUSY GUT: a ``universal seesaw'' mechanism and the higgses
present in the missing VEV mechanism. The universal seesaw is used to
to transfer the mass matrices from a set of superheavy fermions living
near GUT to the ordinary light ones. the higgses in the $\R{45}$
representation that realize the missing VEV mechanism automatically
generate the correct horizontal hierarchies $\e_u\ll\e_{e,d}$.
Then the mass matrices are generated by coupling one of these higgses
with three flavour triplets scalar fields, that define three rank-one
projectors.  This model realization via universal seesaw efficiently
suppresses proton decay and SUSY FCNC, and maintains almost pure MSSM
content below the GUT scale.

In the flavour sector, the solution emerging from the data-fitting
points to quasi-aligned or quasi-planar triplets, that seem to suggest
a precise mechanism of SU(3) breaking along VEVs without introducing
ad-hoc breakings of flavour, as described for example
in~\cite{disoriented,sextets}.

\medskip

We conclude that the unification of Yukawa determinants at the GUT
scale nicely fits into the framework of SO(10) Grand Unification with
horizontal symmetries, where it gives a hint for building the flavour
sector of the theory, while still giving model independent predictions
for the fermion masses and the the $\tb$ MSSM parameter.

\section{Acknowledgments}

This work was partially supported by the MIUR grant under the Projects
of National Interest PRIN 2004 "Astroparticle Physics".

\appendix

\section{Appendix: MSSM renormalization}

\label{app:RG}

In the MSSM, fermion masses and mixing angles can be defined as
follows in terms of quantities at the GUT scale:
\bea
&
\ba[b]{lllll}
m_u=\l_uv\sin\beta\, \eta_{uds} R_u B_t^3 &\quad&
m_d=\l_dv\cos\beta\, \eta_{uds} R_d &\quad&
m_e=\l_ev\cos\beta\, \eta_{e\mu}R_e
\\
m_c=\l_cv\sin\beta\, \eta_c R_u  B_t^3  &\quad&
m_s=\l_sv\cos\beta\, \eta_{uds} R_d  &\quad&
m_\mu=\l_\mu v\cos\beta\, \eta_{e\mu}R_e 
\\
m_t=\l_tv\sin\beta\, \eta_t R_u  B_t^6 &\quad&
m_b=\l_bv\cos\beta\, \eta_b R_d  B_t   &\quad&
m_\tau=\l_\tau v\cos\beta\, \eta_\tau R_e
\ea\\
&
\t_{12}|_Z=\t_{12}
\,,\qquad
\t_{13}|_Z=\t_{12}\,B_t^{-1}
\,,\qquad
\t_{23}|_Z=\t_{12}\,B_t^{-1}\,,
\eea
where the factors $R_{u,d,e}$ account for the running induced by the
MSSM gauge sector, from $M_{GUT}$ to $m_t$; the factor $B_t$ accounts
for the running induced by the large $\l_t$; and $\eta_{e\mu,uds,b,c}$
complete the QCD+QED running from $m_t$ down to 2\GeV\ for u, d, s or
to the respective masses for b, c.

From \cite{BBO},  assuming $\a_3|_Z=0.118$ and $M_{SUSY}=m_t$, we find
\be
\eta_{uds}=1.74\,,\quad \eta_c=2.11\,,\quad \eta_b=1.52\,,\quad
\eta_\tau=1.04\,,\quad \eta_{e\mu}=1.06\,,\quad \eta_t=1\,.
\ee
Then we calculate
\be
R_u=3.45\,,\qquad R_d=3.36\,,\qquad R_e=1.51\,, \qquad R_{\nu_D}=1.39\,,
\ee
Finally, the factor $B_t$ is a function of $\l_t$. We obtain
$B_t|_{\l_t\!=\!0.5}=0.901$, $B_t|_{\l_t\!=\!3.5}=0.670$.  Near
lower $\l_t$, as in the case of moderate or large $\tb$, the
dependence on $\l_t$ is stiffer, and in power 6 it allows a quite
precise dermination of $\l_t$ from the experimental valoue of $m_t$.

The dependence on $\a_3|_Z$ of all these coefficients affects mainly
the lighest quarks via $\eta_{uds}$, $\eta_c$ (10\%), all the others
have uncertainty of the order of 1-3\%. There is a stronger dependence
on the supersymmetry breaking scale.

\end{document}